\documentclass[english,pra, two column]{revtex4-1}
\usepackage[T1]{fontenc}
\usepackage[latin9]{inputenc}
\usepackage{color}
\usepackage{graphicx}

\makeatletter

\newcommand{\lyxdot}{.}

 \@ifundefined{textcolor}{}
 {%
   \definecolor{BLACK}{gray}{0}
   \definecolor{WHITE}{gray}{1}
   \definecolor{RED}{rgb}{1,0,0}
   \definecolor{GREEN}{rgb}{0,1,0}
   \definecolor{BLUE}{rgb}{0,0,1}
   \definecolor{CYAN}{cmyk}{1,0,0,0}
   \definecolor{MAGENTA}{cmyk}{0,1,0,0}
   \definecolor{YELLOW}{cmyk}{0,0,1,0}
 }

\makeatother

\usepackage{babel}
\begin{document}

\title{Entanglement,\textcolor{red}{{} }number fluctuations and optimized
interferometric phase measurement}

\author{Q. Y. He$^{1,2}$}

\author{T. G. Vaughan$^{1,3}$}

\author{P. D. Drummond$^{1}$}

\email{pdrummond@swin.edu.au}

\selectlanguage{english}%

\author{M. D. Reid$^{1}$}

\email{mdreid@swin.edu.au}

\selectlanguage{english}%

\affiliation{$^{1}$Centre for Atom Optics and Ultrafast Spectroscopy, Swinburne
University of Technology, Melbourne 3122, Australia}

\affiliation{$^{2}$State Key Laboratory of Mesoscopic Physics, School of Physics,
Peking University, Beijing 100871, People\textquoteright{}s Republic
of China }

\affiliation{$^{3}$Computer Science Department, Auckland University, Auckland,
New Zealand.}
\begin{abstract}
We derive a phase-entanglement criterion for two bosonic modes which
is immune to number fluctuations, using the generalized Moore-Penrose
inverse to normalize the phase-quadrature operator. We also obtain
a phase-squeezing criterion that is immune to number fluctuations
using similar techniques. These are utilized to obtain an operational
definition of relative phase-measurement sensitivity, via analysis
of phase measurement in interferometry. We show that these measures
are proportional to enhanced phase-measurement sensitivity. The phase-entanglement
criterion is a hallmark for a new type of quantum squeezing, namely
planar quantum squeezing. This has the property that it squeezes two
orthogonal spin directions simultaneously, which is possible owing
to the fact that the SU(2) group that describes spin symmetry has
a three-dimensional parameter space, of higher dimension than the
group for photonic quadratures. The practical advantage of planar
quantum squeezing is that, unlike conventional spin-squeezing, it
allows noise reduction over all phase-angles simultaneously. The application
of this type of squeezing is to quantum measurement of an unknown
phase. We show that a completely unknown phase requires two orthogonal
measurements, and that with planar quantum squeezing it is possible
to reduce the measurement uncertainty independently of the unknown
phase value. This is a different type of squeezing to the usual spin-squeezing
interferometric criterion, which is only applicable when the measured
phase is already known to a good approximation, or can be measured
iteratively. As an example, we calculate the phase-entanglement of
the ground state of a two-well, coupled Bose-Einstein condensate,
similar to recent experiments. This system demonstrates planar squeezing
in both the attractive and repulsive interaction regimes.
\end{abstract}


\maketitle

\section{Introduction}

Entanglement criteria are widely used to identify nonclassical resources
for potential applications in quantum technology. One application
is the enhancement of measurement sensitivity. In practice, the most
sensitive measurements are often interferometric. Hence, the measurement
of an unknown quantity is reduced to the measurement of a phase-shift.
In this communication, we analyze how non-classical, entangled states
can increase phase-measurement sensitivity. To achieve this, we will
introduce both an operational measure of relative phase, and a corresponding
signature of phase entanglement between two Bose fields, using well-defined
interferometric particle-counting procedures. This is shown to quantitatively
measure the enhancement of an interferometric measurement. It is the
interferometric equivalent of the spin-squeezing criterion \cite{spinsqkiti,spinsq,spinsqwine},
which is known to measure the non-classical precision of a clock \cite{spinsqwine}.

We introduce a relative phase operator which is well-defined in the
case of variable total particle number, by using the generalized inverse
method to prevent singularities in the inverse number operator. It
is the simplest relative phase operator that is measurable interferometrically.
Number fluctuations are always found in current experimental photonic
and atom interferometer phase measurements. Hence, we clarify the\textcolor{blue}{{}
}operational phase-measurement procedure already used heuristically
to analyze experiments \cite{Esteve2008,Gross2010,Treutlein,Andrei}.
Most importantly, we show that when the measured phase is unknown
prior to measurement, a state preparation that involves mode-entanglement
is optimal, and is closely related to the relative-phase operator.
This complements previous studies which generally assume either that
the phase is already known to a good approximation, or that the phase-shift
remains constant during repeated measurements. Here, we use the minimal
number of measurements possible. This is inherently different to strategies
employed to estimate phase through sequential or multiple measurements,
which assume the phase-shift is a classical, time-invariant quantity.

Our entanglement measure is a normalized form of the recently introduced
Hillery-Zubairy (HZ) non-hermitian operator product criterion \cite{hillzub},
similar to that introduced in a previous paper \cite{PRL}. We prove
that this normalized form is a phase-entanglement signature for two
Bose fields, and has the advantage of being almost immune to total
number fluctuations. We show how this criterion can be interpreted
as a variance measure which quantifies entanglement, and has a direct
physical interpretation as the enhancement of phase measurement sensitivity
in an interferometer. This is directly related to the idea of planar
quantum squeezing, in which the quantum noise is simultaneously reduced
in two orthogonal directions of phase-measurement. 

The present analysis focuses on linear, two-mode interferometry with
particle-counting detectors. This technique is by far the most commonly
used technique for phase-measurement, and therefore deserving of a
careful analysis. Our results, while less general than an analysis
of completely arbitrary quantum measurements, are able to be implemented
immediately, since two-mode interferometers are widely available both
for photonic and atomic fields. We discuss techniques for generating
the required entangled input fields using atomic sources. As an example,
we consider how to obtain these types of quantum state through the
creation of a correlated ground-state in a coupled, two-mode Bose-Einstein
condensate (BEC) with either attractive or repulsive S-wave scattering
interactions. We show that as well as giving sub-shot noise (squeezed)
phase noise in two orthogonal phase directions simultaneously, it
is possible to obtain nearly Heisenberg-limited peformance in one
of two phase directions, which is useful when the phase is known approximately.

\section{Quantum phase measurements}

Interferometers are designed to measure a relative phase-shift $\phi$,
typically between two beams. The phase information must first be encoded
on quantum fields before it is measured. Hence, there is a close relationship
between interferometry, which measures a phase-shift in a medium,
and the measurement of phase of a quantum field, which is where the
interferometric phase information is stored. The relationship is more
important than meets the eye. An experimentally measured phase-shift
is can neither be truly classical nor time-invariant, which are commonly
used assumptions. For a variety of reasons, repeated measurement is
not always possible, and one must regard interferometric measurement
as primarily a quantum measurement problem. 

In view of this, we first review earlier approaches to phase measurement
of quantum fields, in order to explain and motivate the approach used
in the rest of the paper.

\subsection{Phase operators}

We start with a generic problem in quantum phase measurement: how
can one measure the phase of a quantum field or mode ($\hat{a}$).
Classically, one divides up a field amplitude into intensity ($N_{a}$)
and phase ($\phi_{a}$) by introducing:
\begin{equation}
a=\sqrt{N_{a}}e^{i\phi_{a}}.\label{eq:phase}
\end{equation}
Next, if one measures the complex amplitude $a$, one simply classically
normalizes to obtain the classical phase, $\phi_{a}=-i\ln\left[a/\sqrt{N_{a}}\right]$.
We note that this gedanken-experiment for classical phase measurement
involves measuring both real and imaginary components of the amplitude. 

Quantum studies of this generic problem date back to the early attempt
of Dirac \cite{Dirac} to define a quantum phase operator from a canonical
commutation relation of form $\left[\hat{\phi}_{a},\hat{N_{a}}\right]=-i.$
The underlying mathematical problem is that a hermitian quantum operator
$\hat{\phi}_{a}$ for the phase of a single harmonic oscillator operator
$\hat{a}$ strictly does not exist, which has been the topic of many
previous studies \cite{SussGlog,Louisell,RMP}. This can be seem most
easily from the commutation relations, $\left[\hat{a}\hat{a}^{\dagger}-\hat{a}^{\dagger}\hat{a}\right]=1$.
If phase is hermitian, then $\hat{a}=\sqrt{\hat{N}}\hat{U}$, where
$\hat{U}$ is a unitary operator such that $\hat{U}^{\dagger}\hat{U}=\hat{U}\hat{U}^{\dagger}=1$,
and $\hat{N}=\hat{a}^{\dagger}\hat{a}$. From the commutators, $\hat{N}-1=\hat{U}\hat{N}\hat{U}^{\dagger}$.
This violates unitarity, since a unitary transformation cannot change
an operator's eigenvalues.

This problem does not of course occur classically, and it is at the
root of quantum phase-measurement problems. However, the idea of a
phase-operator with canonical commutators is approximately valid at
large particle number, and suggests the existence of a fundamental
uncertainty principle called the Heisenberg limit:
\[
\Delta\phi\Delta N\ge\frac{1}{2}.
\]
Given that $\Delta N\leq N_{max}/2$, this has led to the idea of
a Heisenberg limit of $\Delta\phi\ge1/N_{max}$ on phase-measurement
with at most $N_{max}$ particles. Sometimes one interchanges $\bar{N}$
and $N_{max}$, as they are often related.

\subsection{Truncated Hilbert space methods}

One resolution of the lack of a hermitian phase operator due to Pegg
and Barnett, is to truncate the Hilbert space to a maximum boson number
of $s$ \cite{PeggBarnett}. Next, one defines a phase eigenstate
$\left|\theta\right\rangle _{p}$ as a discrete Fourier transform
of number states $\left|n\right\rangle $, using:
\begin{equation}
\left|\theta\right\rangle _{p}=\frac{1}{\sqrt{s+1}}\sum_{n=0}^{s}e^{in\theta}\left|n\right\rangle .
\end{equation}

From this starting point, it is clear that a hermitian phase operator
can simply be obtained from the definition:
\begin{equation}
\hat{\phi}_{a}=\sum_{m=0}^{s}\theta_{m}\left|\theta_{m}\right\rangle \left\langle \theta_{m}\right|.
\end{equation}
Here $\theta_{m}=\theta_{0}+2\pi m/\left(s+1\right)$, and $ $$\theta_{0}$
is a reference phase. This is a mathematically consistent approach,
which resolves the issue of hermiticity, but it leaves a number of
practical questions unanswered. 

In particular, what is the physical meaning of the maximum number
$s$ and reference phase $\theta_{0}$? Is it possible to take the
limit of large $s$ in a unique way? From an operational perspective,
what (if any) is the relationship between the abstract operator $\hat{\phi}_{a}$
and an interferometric measurement? One of the purposes of this paper
is to understand how these questions can be answered and implemented
using interferometry.

\subsection{Relative phase operator}

Another solution along these directions is to define a relative phase
operator, $\hat{\phi}=\hat{\phi}_{a}-\hat{\phi}_{b}$, for two modes
$a$ and $b$ \cite{Luis}. This has the conceptual advantage that
it corresponds to operational procedures - which always involve relative
phase measurement - and clarifies the meaning of the reference phase.
A relative phase operator is consistent philosophically with the fundamental
idea of relative measurements in physics, and is the most operationally
meaningful way to define quantum phase in many cases. 

With this approach, one works in a space whose algebra is defined
by the equivalent angular momentum operators in the Schwinger representation:
\begin{eqnarray}
\hat{J}^{X} & = & \frac{1}{2}\left(\hat{a}^{\dagger}\hat{b}+\hat{a}\hat{b}^{\dagger}\right),\nonumber \\
\hat{J}^{Y} & = & \frac{1}{2i}\left(\hat{a}^{\dagger}\hat{b}-\hat{a}\hat{b}^{\dagger}\right),\nonumber \\
\hat{J}^{Z} & = & \frac{1}{2}\left(\hat{a}^{\dagger}\hat{a}-\hat{b}^{\dagger}\hat{b}\right),\nonumber \\
\hat{N} & = & \hat{a}^{\dagger}\hat{a}+\hat{b}^{\dagger}\hat{b}.\label{eq:Angular-Momentum}
\end{eqnarray}

It is usual to assume that one has an eigenstate of the total number
$\hat{N},$ with eigenvalue $N$ and hence an equivalent angular momentum
eigenstate of $J=N/2$. As we show later, this assumption is questionable
in real experimental measurements. The states of well-defined phase
then correspond to linear combinations of angular momentum eigenstates
with $\hat{J}^{Z}|J,m\rangle=m|J,m\rangle$, and one then has a physical
phase basis of:
\begin{equation}
\left|\theta\right\rangle _{p}=\frac{1}{\sqrt{2J+1}}\sum_{m=-J}^{J}e^{im\theta}|J,m\rangle.\label{eq:two-mode phase operator}
\end{equation}

This allows the definition of a relative phase-difference operator
which is Hermitian, and has a discrete spectrum. The fact that the
angular momentum Hilbert space is finite provides a natural explanation
of the truncation parameter $s$ in the Pegg-Barnett approach. However,
it is still not immediately clear how to relate this abstract proposal
to interferometric phase-measurements. 

We also note that in practical interferometry experiments, it is nearly
impossible have an input state that has a well-defined total particle
number, especially at large mean particle number. Instead, the most
common situation is that there is an initial mixture of particle numbers,
with number fluctuations that are typically at least Poissonian. We
will return to the issues of interferometric measurement and number
fluctuations in later sections.

\subsection{Quantum sine and cosine operators}

An alternative resolution of the phase-measurement question is to
define quantum sine and cosine operators. This is a way to reach the
phase through measurement of the real or imaginary part of the amplitude,
an idea which has a clear analog in the classical world, as described
above. Operationally, this is the quantum version of a proposal by
Zernike \cite{Zernike} to analyze coherence in classical interferometry.
The original idea of Zernike was to relate coherence properties directly
to measured classical fringe visibility. 

Extending this to the quantum theory of a single quantized mode \cite{SussGlog},
one can define a normalized amplitude:
\begin{equation}
\hat{E}=\left[1+\hat{N}\right]^{-1/2}\hat{a},
\end{equation}
from which the sine and cosine operators are obtained via:
\begin{eqnarray}
\hat{C} & = & \frac{1}{2}\left[\hat{E}+\hat{E}^{\dagger}\right],\nonumber \\
\hat{S} & = & \frac{1}{2i}\left[\hat{E}-\hat{E}^{\dagger}\right].\label{eq:Quantum_Sine_Cosine}
\end{eqnarray}
This is still operationally unclear for general applications, except
in the limit where one arm of an interferometer is a large `classical-like'
local oscillator. In this limit, the approach is closely related to
the theory of optical quadratures, where the two operators involved
have commutators similar to the quantum position and momentum operators:
for a field mode $\hat{a}$, the quadrature amplitudes are defined
$\hat{X}=(\hat{a}+\hat{a}^{\dagger})/2,$ and\textcolor{red}{{} }\textcolor{black}{$\hat{P}=(\hat{a}-\hat{a}^{\dagger})/\left(2i\right)$. }

A drawback of the local oscillator approach implicit in this method
is that in practical terms it is highly resource-hungry. A classical
local oscillator must necessarily involve a very large boson number
in the local oscillator mode, even though this mode is not formally
included as part of the measured system. We note that there is a generic
issue, which is that since these operators do not commute, it is not
possible to measure both quadratures simultaneously.

\subsection{Relative phase quadrature operators}

Alternatively, the same general idea of quadrature measurement can
be applied to the relative phase between two modes. This also allows
a better understanding of the actual resources involved in the measurement,
since it effectively combines both the measured beam and the local
oscillator into the measured operator. The combined approach is studied
in an early review of Carruthers and Nieto \cite{RMP}, and reduces
to defining the cosine and sine operators of the phase-difference
as:
\begin{eqnarray}
\hat{C}_{12} & = & \hat{C}_{1}\hat{C}_{2}+\hat{S}_{1}\hat{S}_{2},\nonumber \\
\hat{C}_{12} & = & \hat{S}_{1}\hat{C}_{2}-\hat{S}_{2}\hat{C}_{1}.\label{eq:CosinePhase-diff}
\end{eqnarray}

This has the virtue that these operators commute with the total particle
number, $\hat{N}=\hat{N}_{1}+\hat{N}_{2}$. This number is kept finite,
and provides an indication of the true resource needed for the measurement.
A related definition for a two-mode BEC was proposed by Leggett \cite{Leggett},
who suggested defining a phase operator following the approach of
Carruthers and Nieto, except that:
\begin{equation}
\hat{E}=\left[\left(N/2-\hat{J}^{Z}\right)\left(N/2+\hat{J}^{Z}+1\right)\right]^{-1/2}\left(\hat{J}^{X}+i\hat{J}^{Y}\right).\label{eq:LeggettPhaseDiff}
\end{equation}

Here the phase is encoded in the rotations in the $X-Y$ plane, a
convention we will follow in this paper unless otherwise noted. In
these two approaches, the proposed quadrature operators were not analyzed
from the perspective of interferometric fringes and their noise properties.
We will show that the operationally most relevant approach to interferometry
measurements is to use a different normalization to either of the
above suggestions.

\section{Phase operators and interferometry}

\begin{figure}
\begin{centering}
\includegraphics[width=0.8\columnwidth]{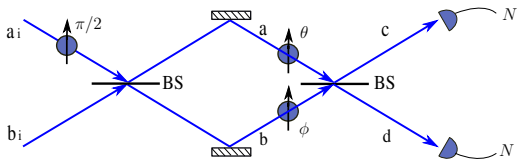}
\par\end{centering}

\caption{{\footnotesize Interferometric measurements can be reduced in the
simplest case to a measurement of outputs from a beamsplitter.} \label{fig:Interferometric-measurements} }
\end{figure}

An operational analysis of interferometric measurements can be reduced
in the simplest case to a measurement of outputs from the final beam-splitter
as shown in figure \ref{fig:Interferometric-measurements}, with one
mode $b$ experiencing an the unknown phase-shift of $\phi$, while
the other mode $a$ is shifted by a fixed reference phase $\theta$.
The other components of the interferometer are then part of the quantum
state preparation that determines the output expectation values. This
is shown schematically as the input beam-splitter with input modes
$a_{i}$, $b_{i}$, in the schematic Mach-Zehnder (MZ) diagram in
figure \ref{fig:Interferometric-measurements}. 

In the following, we shall mostly focus on the general scheme in which
the intermediate modes $a$, $b$ have an arbitrary state preparation.
However, we shall also treat particular examples where the state preparation
is obtained through the Mach-Zehnder protocol. In this case we analyze
the state preparation of modes $a_{i}$, $b_{i}$, as a practical
route towards preparing the intermediate modes, and also introduce
an additional phase-shift on the input to the MZ, for reasons explained
in the last section. 

\subsection{Quantum limits to classical phase-estimation}

The problem of measuring a quantum phase $\hat{\phi}$ is related,
but not identical, to the analysis of quantum limits to estimation
of a classical phase. There is an essential difference, since classical
phase-estimation usually assumes there is a phase-shifting element
that produces a well-defined classical phase-shift $\phi$, which
is supposed to be time-independent. Many treatments assume that the
phase can be measured iteratively without disturbing it, to improve
the accuracy. This limitation rules out many situations where the
phase evolves in time - or where the phase experiences a back-action
which changes the phase after measurement. Other approaches to the
problem make the assumption that it is possible to construct arbitrary
quantum states and measuring devices. As a result, these treatments
generally are not applicable to two-mode interferometry, although
they may be applicable to some future phase-measuring device.

We first introduce earlier approaches to phase-estimation. Pioneering
work by Caves \cite{Caves}, who treated bosons in the context of
gravity-wave detection, showed that two-mode interferometry sensitivity
could be improved above the shot-noise level i.e., the \emph{Standard
Quantum Limit }(SQL) of $\Delta\phi=1/\sqrt{\bar{N}}$: 
\begin{equation}
\Delta\phi<1/\sqrt{\bar{N}}\label{eq:sqph}
\end{equation}
This required non-classical,\textcolor{red}{{} }`squeezed state' input
radiation, reaching a maximum sensitivity near the \emph{Heisenberg
limit} of
\begin{equation}
\Delta\phi=1/\bar{N}\label{eq:hlimit}
\end{equation}
 for an input state with $\bar{N}$ average particle number. Squeezed
states allow the uncertainty of one observable to be reduced below
the standard quantum limit, at the expense of the complementary observables,
so that the Heisenberg uncertainty relation is still satisfied \cite{yuen}.
Thus, for single mode optical amplitudes where $\hat{X}=(\hat{a}+\hat{a}^{\dagger})/2,$
and $\hat{P}=(\hat{a}-\hat{a}^{\dagger})/\left(2i\right)$, for which
$\Delta\hat{X}\Delta\hat{P}\geq1/4$, squeezing of $\hat{X}$ occurs
when $\Delta\hat{X}<1/2$. This is clearly very closely related to
the quadrature phase-operator \cite{SussGlog} approach of equation
(\ref{eq:Quantum_Sine_Cosine}).

Paradoxically, the usual squeezed state technologies of parametric
down-conversion are rather inefficient. The total resources employed,
in terms of boson number prior to down-conversion, are generally no
better than coherent interferometry\cite{Dechoum} for a given phase
sensitivity. As pointed out by Caves, there is still an advantage
of this method for gravity-wave detectors, as it reduces the back-action
caused by radiation pressure.

The treatment of Caves also assumed prior knowledge of the unknown
phase, to allow a linearized treatment in the limit of small fluctuations.
This type of analysis was applied to fermion interferometry by Yurke
\cite{Yurke,ymck}, and was later extended to multiple measurements
\cite{Lane}, with similar conclusions, except for the replacement
of $\bar{N}$ by the total number of particles involved, $\bar{N}_{TOT}$. 

Squeezed quantum fluctuations have been shown to enhance the sensitivity
of other sorts of measurements \cite{tomb,lane2spec}. Wineland et.
al.,  in the context of atomic clock measurements \cite{spinsqwine},
showed that there was a close relationship between interferometry
and the concept of spin-squeezing \cite{spinsqkiti}. The uncertainty
relation for spin is\textcolor{red}{{} }\textcolor{black}{$\Delta\hat{J}^{Y}\Delta\hat{J}^{Z}\geq|\langle\hat{J}^{X}\rangle|/2$},
and spin squeezing exists when the variance of one spin is reduced
below the standard quantum limit (SQL) \cite{spinsqkiti}:
\begin{equation}
\Delta\hat{J}^{Y}<\sqrt{|\langle\hat{J}^{X}\rangle|/2}.\label{eq:spinsq-1}
\end{equation}
The spin squeezing factor 
\begin{equation}
\xi_{S}=\frac{\sqrt{2J}\Delta\hat{J}^{Y}}{|\langle\hat{J}^{X}\rangle|}\label{eq:spinsqfactor}
\end{equation}
was introduced for a collection of $N$ two-level atoms or equivalently,
for two occupied modes for which a collective pseudo-spin is defined,
and $J=N/2$. In spectroscopy or interferometry, the final measurement
is that of a spin component, measured as a population difference between
the two quantum states. 

Since we wish to consider phase in the initial state prior to a beamsplitter,
we consider that the $x$ direction is chosen to be that of the large
spin vector, so that $\langle\hat{J}^{X}\rangle\sim J$. Fluctuations
are thus spin-squeezed when $\Delta^{2}\hat{J}^{Y}<J/2=N/4$. The
precision of the quantum measurement is $ $given by \cite{spinsqwine,Dowling}
\begin{equation}
\Delta\phi=\xi_{S}/\sqrt{N}\label{eq:spsqfint}
\end{equation}
 which is enhanced over the SQL (\ref{eq:sqph}) when
\begin{equation}
\xi_{S}<1\label{eq:spinsq<1}
\end{equation}
 and reaches the Heisenberg limit as $\xi_{S}\rightarrow1/\sqrt{N}.$
This implies that there are large fluctuations in $\hat{J}^{Z}$.
The most extreme case of this is when we choose the input state to
be the eigenstate of the two-mode phase operator, Eq (\ref{eq:two-mode phase operator}),
which establishes a connection between these two approaches. In this
case, one finds that 
\begin{equation}
\Delta^{2}\hat{J}^{Z}=\frac{N}{12}\left[N+2\right].
\end{equation}

Further analysis has suggested ways to reach these Heisenberg limits
(\ref{eq:hlimit}) through macroscopic superpositions or `N00N' states
\cite{Dowling}. More recently, nonlinear interferometry \cite{NonlinearBEC}
was suggested as a route to go beyond the Heisenberg limit, although
this requires specific nonlinear couplings. We note that all of these
techniques use a linearized approach to phase-estimation. This means
that the phase must already be known to an excellent approximation
prior to measurement. In practice, this implies repeated measurements
to refine the estimation of the phase until the final, high-precision
measurement is made. In particular, we show below that a two-mode
phase eigenstate only gives a low interferometric measurement variance
when the phase is known almost perfectly prior to the measurement.

There are general treatments that typically involve more than single-pass,
two-mode interferometry \cite{Braunstein}. Sanders and Milburn \cite{sandersmilburn}
found the optimal measurement and state to determine the phase $\phi$,
based on the two-mode phase-operator approach of equation (\ref{eq:two-mode phase operator}).
Berry and Wiseman \cite{berrywise,BerryWiseman} showed that this
canonical measurement cannot be realized by counting particles in
an interferometer (figure \ref{fig:Interferometric-measurements}),
and proposed alternative iterative schemes. Such idealized measurements
have also been analyzed by many authors using techniques like the
quantum Fisher information \cite{fisher,Benatti}. 

The assumption of a time-invariant, classical phase-shift used in
iterative schemes can rule out such techniques for many applications.
Phases often vary in time, and one cannot always avoid quantum back-action.
The physical reason for this is that a phase-shift corresponds to
an energy-shift term in the radiation-matter Hamiltonian \cite{dynamical-renormalization}.
This therefore has dynamical consequences for the object being measured,
and may change the phase. Hence, the use of iterative or repeated
measurement schemes is not always feasible. Strategies of this type
are different to the present theory, which focuses on minimal numbers
of measurements using a two-mode linear interferometer.

\subsection{Phase-measurement operator}

We now return to the fundamental question of how to measure a phase-shift
using a quantum phase-measurement operator. We wish to use a strictly
operational definition, solely utilizing the interferometer outputs.
If we define the operator phases so that $c,d=\left[ae^{-i\theta}\pm be^{-i\phi}\right]/\sqrt{2}$,
then the measured outputs of the quantum interferometer in terms of
$a,b$ are:
\begin{eqnarray}
\hat{N}_{\pm} & = & \hat{c}^{\dagger}c\pm\hat{d}^{\dagger}d\nonumber \\
 & = & \frac{1}{2}\hat{N}\pm\frac{1}{2}\left[\hat{a}^{\dagger}\hat{b}e^{-i(\phi-\theta)}+\hat{b}^{\dagger}\hat{a}e^{i(\phi-\theta)}\right].\label{eq:numberdiff}
\end{eqnarray}
Comparing these quantities with the equivalent angular momentum operator
approach from equation (\ref{eq:Angular-Momentum}), we see that these
quantities can be rewritten as: $\hat{N}_{\pm}=\hat{N}/2\pm\hat{J}^{\phi}$,
where $\hat{N}$ is the total number operator, and:
\[
\hat{J}^{\phi}=\hat{J}^{X}\cos(\phi-\theta)+\hat{J}^{Y}\sin(\phi-\theta).
\]
 In any given measurement, there are always two outputs which can
be measured simultaneously. The quantum phase of the output field
- which is also related to the unknown phase-shift $\phi$ - can be
estimated from the ratio:
\begin{eqnarray}
\tilde{J}\left(\phi\right) & = & \lim_{\epsilon\rightarrow0+}\frac{\hat{N}\left(\hat{N}_{+}-\hat{N}_{-}\right)}{2\left(\epsilon+\hat{N}^{2}\right)}=\hat{J}^{\phi}\hat{N}^{+}.\label{eq:numberdiffnorm}
\end{eqnarray}
Here $\hat{N}^{+}$ is the Moore-Penrose generalized inverse \cite{Moore-Penrose}
of $\hat{N}$. This is a well-defined hermitian observable that commutes
with $\hat{J}^{\phi}$, and gives the least-squares solution of any
inversion or variational problem involving $\hat{N}$. The generalized
inverse $\hat{N}^{+}$ has many of the properties of a standard inverse,
including the property that its eigenvalues are $N^{-1}$ for number
states with total number $N>0$. However, it is zero (not infinity)
for the vacuum state. This means that, unlike the standard inverse,
it has a well-defined value for \emph{all} quantum states. More details
are in the Appendix. 

In experiments designed for accurate phase measurement for large average
particle numbers, events with zero total particle number occur with
vanishingly small probability, and generally can be neglected, in
which case $\hat{N}^{+}\approx\hat{N}^{-1}$. Of course, this is only
true approximately, as the number operator has no standard inverse.
The quantity $\tilde{J}\left(\phi\right)$ is always measurable and
hermitian. Hence, $\tilde{J}\left(\phi\right)$ can be called a relative
phase quadrature operator, and is the fundamental quantity measured
in any phase-sensitive interferometric experiment. 

It is vital to normalize by the particle number at each measurement
- as indicated in the above operator - for the simple reason that
in general, the particle number is not known in advance. It is often
theoretically assumed that the total particle number is known prior
to measurement. This is rarely found in real experiments, especially
as the number is increased. An inspection of the experimental protocols
actually used in recent BEC interferometry experiments \cite{Esteve2008,Treutlein,Andrei}
shows that the operator given above corresponds rather closely to
the way that data is analyzed in practice. Our analysis therefore
provides a theoretical justification for these operational procedures.
We will treat the effects of typical Poissonian particle number fluctuations
in a later section.

The operator $\tilde{J}\left(\phi\right)$ is different to both the
complex phase-difference quadrature operators of Nieto and Carruthers
\cite{RMP}, and of Leggett, owing to the type of normalization chosen
here. If we compare the current approach of equation (\ref{eq:numberdiffnorm})
with these earlier suggestions in equations (\ref{eq:CosinePhase-diff})
and (\ref{eq:LeggettPhaseDiff}), there is a very important difference.
The operator $\tilde{J}\left(\phi\right)$ given above is uniquely
defined for all inputs, and can be measured completely from a single,
combined measurement of the two interferometer outputs. It is not
obvious how one can measure the earlier proposed phase-measurement
operators in practical interferometry experiments, since they appear
to require the simultaneous measurement of non-commuting output operators
like $\hat{J}^{X}$ and $\hat{J}^{Z}$. Of course, this does not rule
out more sophisticated operational measurements, as the combined operators
are hermitian; but these measurements do not appear feasible with
simple beam-splitters and photodetectors. 

If multiple measurements are made sequentially, then more sophisticated
iterative phase-estimation techniques are possible \cite{Lane,Braunstein,berrywise,recentparity,griffithexp}.
As pointed out above, this does not help in experiments where the
phase is changing. Often, only a single measurement is possible\textcolor{black}{.
W}e also recall that many previous analyses are conditioned on having
\emph{a priori }approximate knowledge of the unknown phase-shift $\phi$.
In the following, we focus instead on optimizing the sensitivity of
the operational phase measurement equation (\ref{eq:numberdiffnorm}),
for a range of unknown phase-shifts. In other words, we assume that
the phase is known to lie in a given interval that is \emph{not} vanishingly
small.

\subsection{Entanglement and squeezing }

Interferometric sensitivity and particle entanglement have previously
been linked through criteria involving Fisher information \cite{fisher}.
Sorenson et. al. \cite{naturespinsqent} have shown that a measure
of particle entanglement is the spin squeezing criterion (\ref{eq:spinsq<1}):
a fixed number $N$ of two-level systems (spin-$1/2$ particles) are
separable when $\rho=\sum_{R}P_{R}\rho_{1}^{R}...\rho_{N}^{R}$ and
hence entangled when 
\begin{eqnarray}
0<\xi_{S}^{Z/Y}=\frac{\sqrt{N}\Delta\hat{J}^{Z/Y}}{|\langle\hat{J}^{X}\rangle|} & < & 1.\label{eq:spinsq}
\end{eqnarray}
Here we\textcolor{black}{{} define the collective spins associated with
$N$ spin $1/2$ systems: $\hat{J}^{\theta}=\sum_{k=1}^{N}\hat{J}_{k}^{\theta}$
($\theta=X,Y,Z$), and $\hat{J}_{k}^{\theta}$ is the spin of the
$k$-th particle. The Heisenberg Uncertainty Principle places a lower
bound on $\xi_{S}$, because of the finite size of the Hilbert space.
The precise values for the lower bound for fixed $N$ have been determined
by Sorenson and Molmer in Ref. \cite{sorenson}, and decrease with
increasing spin $J$. The spin squeezing criterion has been measured
experimentally in BEC interferometry ~\cite{Esteve2008,Gross2010,Treutlein},
and is related to phase-measurement efficiency when the phase value
is approximately known in advance \cite{spinsqwine}, as summarized
by (\ref{eq:spsqfint}). }

\textcolor{black}{In this paper, we will generalize this criterion
to include number fluctuations, and also treat a very different type
of entanglement, that between two distinct spatially separated locations,
rather than between many qubits. We will relate the sensitivity of
the phase measurement (\ref{eq:numberdiffnorm}) of an unknown phase
to a special type of two-mode entanglement between $a$ and $b$.
Two-mode entanglement is defined as a failure of the separable model
\begin{equation}
\rho=\sum_{R}P_{R}\rho_{a}^{R}\rho_{b}^{R}\label{eq:twomodeent}
\end{equation}
where $P_{R}>0$, $\sum_{R}P_{R}=1$ and $\rho_{a/b}^{R}$ are density
operators for states at $a/b$. }

\textcolor{black}{Many criteria for two-mode entanglement exist, but
we are interested only in those interferometric measures of entanglement
that can enhance the phase measurement task of Figure \ref{fig:Interferometric-measurements}
\cite{holburn,dowlent,fisher}. The observables that can be measured
are given in equation (\ref{eq:Angular-Momentum}) and (\ref{eq:numberdiffnorm}).
These expressions are written in terms of the modes $a$ and $b$
that are the inputs to the final beam splitter. One may also consider
a MZ type of experiment with a phase-rotation and two beam-splitters
as is illustrated in Figure \ref{fig:Interferometric-measurements}.
In this case, the MZ internal modes are related to the input modes
$a_{i},b_{i}$ by $a=(-ia_{i}+b_{i})/\sqrt{2}$, $b=-(ia_{i}+b_{i})/\sqrt{2}$,
and the measured outputs are simply rotated versions of the Mach-Zehnder
input operators, with:
\begin{eqnarray}
\hat{J}^{X} & = & \hat{J}_{i}^{Z},\nonumber \\
\hat{J}^{Y} & = & -\hat{J}_{i}^{X}.
\end{eqnarray}
}

\textcolor{black}{We see from these expressions (\ref{eq:Angular-Momentum})
and (\ref{eq:numberdiffnorm}) that the sensitivity of the phase measurement
will depend on the noise levels of the two orthogonal components,
$\hat{J}^{X}\hat{N}^{+}$ and $\hat{J}^{Y}\hat{N}^{+}$ (in terms
of the interferometer modes $a$ and $b$). It makes sense to then
choose an input state for the interferometer that will maximally reduce
the noise in }\textcolor{black}{\emph{both}}\textcolor{black}{{} of
these components simultaneously. In fact this requirement is closely
related to an entanglement measure. Hillery and Zubairy (HZ) showed
\cite{hillzub} that for any }\textcolor{black}{\emph{separable }}\textcolor{black}{state
(\ref{eq:twomodeent}), 
\begin{equation}
\Delta^{2}\hat{J}^{X}+\Delta^{2}\hat{J}^{Y}\geq\langle\hat{N}\rangle/2.\label{eq:hzsep}
\end{equation}
Entanglement between modes $a$ and $b$ is thus detected when (\ref{eq:hzsep})
fails: 
\begin{eqnarray}
0 & < & E_{HZ}=\frac{\Delta^{2}\hat{J}^{X}+\Delta^{2}\hat{J}^{Y}}{\langle\hat{N}\rangle/2}<1.\label{eq:hz}
\end{eqnarray}
$E_{HZ}=1$ gives the Standard Quantum Limit (SQL) noise level, which
is the noise level $E_{HZ}$ obtained when the modes $a$ and $b$
are in the separable product of coherent states, $|\alpha\rangle|\beta\rangle$.
It is not possible however to choose a state so that both variances
$\Delta^{2}\hat{J^{X}}$, $\Delta^{2}\hat{J^{Y}}$ are zero.}

\subsection{\textcolor{black}{Planar Quantum Squeezing\label{sub:Planar-Quantum-Squeezing}}}

\textcolor{black}{If we consider the Heisenberg uncertainty principle
in the $X-Y$ plane, we see that it has the form $\Delta\hat{J}^{Y}\Delta\hat{J}^{X}\geq|\langle\hat{J}^{Z}\rangle|/2$.
Here the optimal situation is obtained for equal beam intensities
entering the beam-splitter, so that $\langle\hat{J}^{Z}\rangle=0$.
This appears to provide no lower bound to the measured quadrature
variances, and hence to the phase noise. However, appearances can
be very misleading. In fact, a non-zero lower bound to $E_{HZ}$ exists,
because the variances of $\hat{J}^{X}$ and $\hat{J}^{Y}$ cannot
be simultaneously zero. }

\textcolor{black}{For fixed $N=2J$, this bound has been determined.
It is known that 
\begin{equation}
C_{J}/J\leq E_{HZ},\label{eq:cj}
\end{equation}
where the coefficients $C_{J}\sim3\left(2J\right)^{2/3}/8$ as $J\rightarrow\infty$
\cite{c_j}. This means, however, that both the orthogonal variances
in a phase-measurement can be simultaneously reduced below the shot-noise
level, since we are minimizing the sum of the phase variances, $\Delta^{2}\hat{J}^{X}+\Delta^{2}\hat{J}^{Y}$.
In general, a noise reduction of the sum of two variances below the
shot-noise level is called planar quantum squeezing or PQS \cite{c_j},
as it minimizes quantum noise in a plane, rather than just in one
direction on the Bloch sphere. It has the advantage that noise reduction
for phase measurement occurs regardless of the value of the unknown
phase.}

\textcolor{black}{It is instructive to compare the optimal PQS state
with the relative phase eigenstate. The minimal variance PQS state
is:
\begin{equation}
|\psi\rangle=\frac{1}{\sqrt{2J+1}}\sum_{m=-J}^{J}R_{m}e^{im\theta}|J,m\rangle.\label{eq:purestate}
\end{equation}
}

\textcolor{black}{The asymptotic limit of the optimal coefficient
$R_{m}$ , which minimizes the sum of the quadrature variances, is
then a Gaussian of form:
\begin{equation}
R_{m}=\frac{e^{-m^{2}/\left(2\sigma_{m}\right)}}{\sqrt{2\pi\sigma_{m}}}.
\end{equation}
where the variance in the space of $\hat{J}^{Z}$ eigenvalues is $\sigma_{m}=\Delta^{2}\hat{J}^{Z}=\left(J^{2}/2\right)^{2/3}$.
The other important properties are that, to leading order: $\langle\hat{J}^{X}\rangle\sim J,\Delta^{2}J^{X}\sim\left(2J\right)^{2/3}/8,\Delta^{2}J^{Y}\sim\left(2J\right)^{2/3}/4$}
. Thus, the optimal PQS state reduces the variance in $ $$J^{X}$
and $J^{Y}$ simultaneously, with \emph{both} variances well below
the shot-noise level. 

By contrast, for a relative phase state at large $J$, we have $\langle\hat{J}^{X}\rangle\sim\pi J/4$,
$\Delta^{2}\hat{J}^{X}\sim(2/3-\pi^{2}/16)J^{2}$, $\Delta^{2}\hat{J}^{Y}\sim\sqrt{\ln\left(J\right)}$,
$\Delta^{2}\hat{J}^{Z}\sim J^{2}/3$. Hence, the Heisenberg uncertainty
principle in the Z-Y plane is obeyed, since $\Delta\hat{J}^{Y}\Delta\hat{J}^{Z}>|\langle\hat{J}^{X}\rangle|/2$$ $.
However, quantum squeezing only occurs in the $Y$ spin direction,
while in both the other spin directions the noise is greatly increased
\emph{above} the shot-noise level. This means that in an interferometric
measurement using a relative phase state, the reference phase-offset
$\theta$ must be adjusted to match $\phi$ with high precision, even
though $\phi$ is of course unknown prior to measurement. This adjustment
is necessary to avoid contamination of the results with high levels
of noise from the $X$ spin direction, which are well above the shot-noise
or Poissonian level. The underlying cause is that interferometric
measurements do \emph{not} simply project out the relative phase eigenstates.

We see that the main advantage of PQS states in interferometry is
that it is possible to have sub-shot precision in both the measured
spin directions simultaneously. This is advantageous when the measured
phase is truly unknown. At first, it may seem that this is less than
optimal as a squeezing strategy when the phase is known approximately.
For the optimal PQS state described above, which minimizes the variance
sum, neither of the variances are close to the Heisenberg limit. Importantly,
there are a range of possible PQS states in which the relative variances
in the $X$ and $Y$ directions can be adapted to the desired measurement
strategy, including states in which PQS - with the variance sum below
the shot-noise level - is combined with nearly Heisenberg-limited
variance reduction in one of the two directions. This possibility
is discussed in the last section, together with practical techniques
for achieving it.

\section{Operational criteria and number fluctuations}

In practical interferometry, the total number of input bosons usually
changes at each measurement. Hence the ensemble used for averaging
has a finite distribution over the particle number. This is caused
by a number of factors. In optical lasers, it is caused by technical
noise in the optical pumping process, as well as well-known quantum
noise effects during stimulated emission and out-coupling \cite{Louisell-book}.
In BEC and atom lasers, the factors involved range from fluctuations
in the initial atomic density distribution in the magneto-optical
trap, to quantum noise due to the atomic collisions that occur in
the evaporative cooling process \cite{EvapCooling}. These number
fluctuations are due to the non-equilibrium mechanisms that generate
a laser or BEC respectively, and there is no reason to assume either
a canonical or a grand canonical ensemble.

The direct use of the HZ spatial entanglement criterion is highly
sensitive to total number fluctuations.\textcolor{red}{{} }For this
reason, we will introduce entanglement and spin-squeezing definitions
that are normalized by the total number. We refer to such normalized
entanglement measures as \emph{phase-entanglement} and \emph{phase-squeezing}
measures, as they measure correlated and reduced noise phases. There
is another possible strategy, which is to simply reject all measurements
that have the `wrong' particle number. This allows a conditional number
state measurement to be obtained. While this is feasible, it is also
extremely inefficient, since most attempted measurements yield no
information at all about the phase with this strategy.

\subsection{Experimental number fluctuations\label{sub:Experimental-number-fluctuations}}

In the case where the number fluctuations are Poissonian, the probability
that there are exactly $N$ bosons is
\begin{equation}
P(N)=\frac{1}{N!}\langle\hat{N}\rangle^{N}e^{(-\langle\hat{N}\rangle)}.
\end{equation}
While this may not be the best input state for a phase measurement,
this distribution does give a number standard deviation of $\sigma_{N}=\sqrt{\bar{N}}$,
which is a typical order of magnitude for the number fluctuations
in a well-stabilized photonic laser or BEC. This fluctuation is in
the total number $N$, prior to any beam-splitter. Interferometric
beam-splitters introduce relative number fluctuations in addition
to the total number fluctuation.

Highly-stabilized semiconductor lasers have reached slightly lower
number variances than Poissonian, in a restricted frequency range
\cite{NumberSqueezing}. For an atomic BEC experiment, atom number
statistics are difficult to measure accurately to this level of precision
for large $\bar{N}$. Number fluctuations of at least the Poissonian
level are found in almost all current BEC experiments \cite{Treutlein}
where data is available. Just as with lasers, it is possible to obtain
lower variances than this, with some restrictions. In the best results
to date, standard deviations as low as $0.6\sqrt{\bar{N}}$ (below
the Poissonian level) were observed at very small atom numbers of
$\bar{N}\approx60$. Super-Poissonian variances were found for larger
numbers of $N>500$ \cite{RaizenBECNumberSqueezing}.

When calculating the entanglement parameter $E_{HZ}$ for typical
states with Poissonian fluctuations, we find that the apparent entanglement
is greatly decreased due to the effect of total number fluctuations.
This is misleading: total number fluctuations do not destroy entanglement.
Accordingly, it is important to use entanglement and phase-sensitivity
measures that allow for number fluctuations. Our general criteria
therefore include number fluctuations with an arbitrary variance.
These criteria can still be used to describe idealized experiments
without number fluctuations, even though this is not very realistic.
In the next section, we will make use of a Poissonian distribution
to model the behavior of typical BEC experiments, with a low atom
number of $\bar{N}\sim100$.

\subsection{Entanglement and squeezing criterion}

To treat phase measurement including total number fluctuations, we
have introduced normalized spin operators: $\tilde{J}^{\theta}=\hat{J}^{\theta}\hat{N}^{+}$.
Here, $\hat{N}^{+}$ is the Moore-Penrose generalized inverse of the
number operator. Detailed properties and proofs are given in the Appendix.
We can now use these normalized operators to derive general operational
criteria for entanglement and squeezing, which extend the results
obtained above to a realistic environment with number-fluctuations.
We note that somewhat different results have been obtained previously
with number fluctuations, in work that used un-normalized operators
\cite{EntanglementwithNumberFluctuations}.

\subsubsection{Phase-entanglement criterion}

We now introduce a phase-entanglement measure that is robust against
number fluctuations. We show in the Appendix that in a number-fluctuating
environment, entanglement between modes $a$ and $b$ is confirmed
via a phase-entanglement criterion that uses the generalized Moore-Penrose
inverse of the number operator, $\hat{N}^{+}$:
\begin{eqnarray}
E_{ph} & = & \frac{\Delta^{2}\tilde{J}^{X}+\Delta^{2}\tilde{J}^{Y}}{\langle\hat{N}^{+}\rangle/2}<1.\label{eq:hznorm}
\end{eqnarray}

\subsubsection{Phase-squeezing criterion}

Similarly, we show in the Appendix that entanglement between $N$
spin-$1/2$ systems is confirmed by a normalized spin squeezing criterion,
which we term phase-squeezing:\textcolor{black}{
\begin{equation}
\xi_{S,ph}^{Z/Y}=\frac{\sqrt{\langle\hat{N}\rangle}\Delta\tilde{J}^{Z,Y}}{|\langle\tilde{J}^{X}\rangle|}<1.\label{eq:spinsqnorm}
\end{equation}
The two criteria (\ref{eq:hznorm}) and (\ref{eq:spinsqnorm}) and
their application to determine the enhanced sensitivity of a two-mode
atom interferometer, in particular a BEC atom interferometer for which
incoming number fluctuations are included, form the major results
of this paper.}

\subsection{\textcolor{black}{Phase sensitivity}}

\textcolor{black}{Next, we will obtain a detailed understanding of
the relationship between our phase-entanglement measure, and phase-measurement
sensitivity. The crucial issue in phase measurement is the measurement
sensitivity, or smallest measurable phase-shift. This is related to
the differential signal to noise ratio, given by \cite{Yurke,ymck}
\begin{equation}
\frac{dS}{d\phi}\equiv(\Delta\phi)^{-1}=\frac{1}{\sqrt{\bigl(\Delta\tilde{J}\bigr)^{2}}}|\frac{d\langle\tilde{J}\rangle}{d\phi}|\,.\label{eq:ratioint}
\end{equation}
The smallest measurable change in phase in a single measurement is
$\Delta\phi$. Figure \ref{fig:Interferometric-measurements} depicts
an unknown phase shift $\phi$ (to be measured) relative to a fixed
phase shift $\theta$. We suppose for simplicity that $\langle\hat{a}^{\dagger}\hat{b}\hat{N}^{+}\rangle=|\langle\hat{a}^{\dagger}\hat{b}\hat{N}^{+}\rangle|$,
so that the direction of the mean spin of the state to be used in
the interferometer will be along the $X$ axis: i.e. when $\phi=0$:
$\langle\hat{J}^{X}\hat{N}^{+}\rangle=|\langle\hat{a}^{\dagger}\hat{b}\hat{N}^{+}\rangle|$,
$\langle\hat{J}^{Y}\hat{N}^{+}\rangle=\langle\hat{J}^{Z}\hat{N}^{+}\rangle=0.$
For a controlled reference phase shift $\theta'$, two successive
orthogonal measurement settings, $\theta'$ and $\theta'+\pi/2$,
will allow a determination of the unknown phase $\phi$:
\begin{eqnarray}
\langle\tilde{J}(\phi,\theta')\rangle & = & {\color{red}{\color{black}\cos(\phi-\theta')}}|\langle\hat{a}^{\dagger}\hat{b}\hat{N}^{+}\rangle|,\label{eq:jint}\\
\langle\tilde{J}\left(\phi,\theta'+\pi/2\right)\rangle & = & -\sin(\phi-\theta')|\langle\hat{a}^{\dagger}\hat{b}\hat{N}^{+}\rangle|.\nonumber 
\end{eqnarray}
}

\textcolor{black}{A single measurement setting $\theta'$ cannot determine
the unknown phase completely, since the information given is about
$\cos\varphi$ only. The mean differential signal for measurement
$\tilde{J}(\phi,\theta')=\hat{J}^{\phi}\hat{N}^{+}$ is $-\langle\hat{J}^{X}\hat{N}^{+}\rangle\sin\varphi$,
$ $and $\Delta\phi$ as given by (\ref{eq:ratioint}) for this measurement
is ($\varphi=\phi-\theta')$ 
\begin{eqnarray}
(\Delta\phi)^{2} & = & \Bigl\{\bigl(\Delta\tilde{J}^{X}\bigr)^{2}\,\cot^{2}(\varphi)+\bigl(\Delta\tilde{J}^{Y}\bigr)^{2}\Bigr\}/\bigl|\langle\tilde{J}^{X}\rangle\bigr|^{2},\label{eq:intexp}
\end{eqnarray}
together with a similar expression obtained in the orthogonal direction.
The objective is to determine the conditions on the interferometric
state so that the uncertainty in the phase estimation is minimized.}

\textcolor{black}{The Standard Quantum Limit (SQL) sensitivity $\Delta\phi=1/\sqrt{N}$,
as given by equation (\ref{eq:sqph}), is obtained when fields $a$
and $b$ are formed from a number state $|N\rangle$ incident at one
port of a beam splitter, with a vacuum state input at the second port
\cite{Yurke,dowlent,Gross2010}. An entangled state results \cite{bec ground state ent},
for which $\langle\hat{J}^{X}\rangle=N/2$, $\bigl(\Delta\hat{J}^{X}\bigr)^{2}=0$
and $\bigl(\Delta\hat{J}^{Y}\bigr)^{2}=N/4$. Then, for all phases
$\varphi$, it is readily shown that $\Delta\phi\sim1/\sqrt{N}$. }

\textcolor{black}{For some entangled states, it is well-known \cite{holburn,multiphoton}
that the phase sensitivity can be enhanced below the SQL. The most
well-studied cases however consider a small phase shift about a fixed
phase reference \cite{spinsqwine,dowlent}. It is evident from (\ref{eq:intexp})
that the maximum differential for $ $$\tilde{J}(\phi,\theta')$ is
at $\varphi=\pi/2$, for which $(\Delta\phi){}_{\pi/2}=\Delta\tilde{J}^{Y}/|\langle\tilde{J}^{X}\rangle|$.
The sensitivity at this point is thus given by the normalized spin
squeezing parameter (\ref{eq:spinsqnorm}) which reduces to (\ref{eq:spinsq})
for fixed number inputs. Sub-shot noise sensitivity is achieved when
$\Delta\phi<1/\sqrt{\langle\hat{N}\rangle}$, so by the definition
(\ref{eq:spinsqnorm}), sub-shot noise enhancement occurs for interferometric
states satisfying 
\begin{equation}
\xi_{S,norm}<1.\label{eq:sqnormintsql}
\end{equation}
The technique relies on an accurate estimate $\phi_{X}$ of the unknown
phase, combined with setting $\theta$ to $\phi_{X}-\pi/2$, so that
subsequent measurements measure small shifts near the optimal phase.
We will show in the next section that a near Heisenberg-limited sensitivity
of $(\Delta\phi)_{\pi/2}\sim O(\sqrt{2}/N)$ is predicted for this
case, when the two-modes $\hat{a}$ and $\hat{b}$ of an atom interferometer
are prepared from a two-mode double-potential well BEC ground state.}

\subsection{Estimation of an unknown phase }

\textcolor{black}{The question of phase estimation with an unknown
phase and limited number of measurements is a different issue \cite{sandersmilburn}.
Where we restrict to phase estimation via the interferometric scheme
figure \ref{fig:Interferometric-measurements} based on the number
difference measurements (\ref{eq:numberdiff}-\ref{eq:numberdiffnorm}),
we see from (\ref{eq:intexp}) that a noise-reduction enhancement
over a range of angles}\textcolor{black}{\emph{, }}\textcolor{black}{with
a reduced variance in}\textcolor{black}{\emph{ both}}\textcolor{black}{{}
$\Delta\hat{J}^{X}$ and $\Delta\hat{J}^{Y}$ is needed. This is essential
where there is no prior knowledge of the phase $\phi$ and successive
adaptive phase measurements \cite{berrywise} are not possible. We
note that the sensitivity of the measurement $ $$\tilde{J}(\phi,\theta')$
is destroyed by the divergent contribution evident in (\ref{eq:intexp})
at $\varphi\sim0,\pi$, unless $\Delta\hat{J}^{X}=0$, which places
a severe limit on the interferometric state. This is not a necessary
consideration, however, if the full phase is to be measured via both
orthogonal measurements settings given by (\ref{eq:jint}). The settings
in the ``quiet'' quadrants $\theta'=(\phi+\pi/2)\pm\pi/4$ and $\theta'=(\phi-\pi/2)\pm\pi/4$
have enhanced sensitivity over those in the ``noisy'' quadrants
$\theta'=\phi\pm\pi/4$ and $\theta'=(\phi+\pi)\pm\pi/4$, and for
any unknown phase $\phi$ one of the orthogonal settings $\theta'$
or $\theta'+\pi/2$ will be in the useful quadrants. Least squares
estimation is an obvious strategy here.}

\textcolor{black}{We thus consider the following strategy. The first
reading of the pair $\tilde{J}(\phi,0)$ or $ $$\tilde{J}(\phi,\pi/2)$
determines values for $\cos\phi$ and $\sin\phi$, and thus the location
of the phase in the plane. In this way, it can be determined which
of $ $$\tilde{J}(\phi,0)$ or $ $$\tilde{J}(\phi,\pi/2)$ has measured
in the quiet quadrants. Sub-shot noise sensitivity is then guaranteed
}\textcolor{black}{\emph{at all unknown angles $\phi$,}}\textcolor{black}{{}
for this preferred measurement, provided it can be shown that $\Delta\phi<1/\sqrt{\langle\hat{N}\rangle}$
across the entire two quiet quadrants. According to (\ref{eq:intexp}),
the worse-case sensitivity for these quiet quadrants is at $\varphi=\pm\pi/4$,
$\pm3\pi/4$, and is given by $(\Delta\phi)_{max}^{2}=\left[\Delta^{2}(\tilde{J^{X}})+\Delta^{2}(\tilde{J^{Y}})\right]/\bigl|\langle\tilde{J^{X}}\rangle\bigr|^{2}$.
The condition for $(\Delta\phi)_{max}^{2}$ to be sub-shot noise is
$(\Delta\phi)_{w}<1/\sqrt{\langle\hat{N}\rangle}$ which is quantified
by a phase-sensitivity measure:
\begin{equation}
\eta_{ph}=\frac{\sqrt{\langle\hat{N}\rangle\langle\hat{N}^{+}\rangle E_{ph}/2}}{|\langle\tilde{J}^{X}\rangle|}<1.\label{eq:phmeasuresubshot}
\end{equation}
When the interferometric fields satisfy (\ref{eq:phmeasuresubshot}),
sub-shot noise sensitivity for all angles $\phi$ is guaranteed, for
any fixed measurement setting $\theta$ within the two quiet quadrants
for measurement of $\phi$. }

\textcolor{black}{The fundamental quantum limit for (\ref{eq:phmeasuresubshot})
is given by the smallness of $\eta_{ph}$, which is linked to the
uncertainty relation for the sum of the two spins $J^{X}$ and $J^{Y}$.
It is therefore important to determine a tight lower bound on this
sum, in order to obtain the ultimate phase interferometric sensitivity.
The real question becomes to what extent can we still minimize $\Delta^{2}\tilde{J^{Y}}$,
given that the sum $\Delta^{2}(\tilde{J^{X}})+\Delta^{2}(\tilde{J^{Y}})$
is also to be minimized. The answer is not the same as for two complementary
observables like two optical quadratures, or position and momentum,
for which the commutator is a constant. The uncertainty relation for
spin operators has a state-dependent form,
\[
\Delta J^{X}\Delta J^{Y}\geq\left|\langle J^{Z}\rangle\right|/2.
\]
which means the two variances $\Delta J^{X}$ and $\Delta J^{Y}$can
both be reduced below the shot noise level. For $\langle J^{Z}\rangle=0$
the Heisenberg uncertainty principle is unable to give any bound at
all. This is possibly misleading, since the Heisenberg uncertainty
principle is simply a bound that does not guarantee it can be saturated.
As discussed in section \ref{sub:Planar-Quantum-Squeezing}, a recent
analysis of the spin-variance uncertainty leads to a tight bound on
the variances:
\[
\Delta^{2}J^{X}+\Delta^{2}J^{Y}\geq C_{J},
\]
where $C_{J}$ is a function of the total spin $J$ with an asymptotic
limit of $3(2J)^{2/3}/8$. For $J=N/2$ one finds $\Delta\phi\geq\sqrt{C_{J}}/J\rightarrow O(\sqrt{1.5}/N^{2/3})$,
which we will show in the next section is predicted for the two-well
BEC ground state, in the both attractive and repulsive regimes. The
fundamental limit is below the SQL of $O(1/\sqrt{N})$ over all the
phase angles in the half-plane, but can only reach the Heisenberg
limit of $O(1/N)$ over part of the range. We next present the details
of how these levels of sensitivity can be realized in a BEC interferometer.}

\section{\textcolor{black}{BEC interferometer}}

\textcolor{black}{We consider how the criteria (\ref{eq:hznorm}),
(\ref{eq:spinsqnorm}) and (\ref{eq:phmeasuresubshot}) for entanglement
and phase-measurement sensitivity are satisfied in a typical cold-atom
experiment. The mechanism for two-mode squeezing and entanglement
here is similar to that first realized for optical modes using four-wave
mixing \cite{ReidWalls}, except employing the ground-state of an
interacting BEC. We consider an idealized two-mode or two-well BEC
with a normalized self-interaction coefficient $g$, and a linear
tunneling coupling $\kappa$ between two modes with boson operators
$\hat{a}$ and $\hat{b}$, described by the Hamiltonian \cite{twomodewellbec}
\begin{equation}
H=\kappa(a^{\dagger}b+ab^{\dagger})+\frac{g}{2}[a^{\dagger}a^{\dagger}aa+b^{\dagger}b^{\dagger}bb].\label{hamgs-1-1}
\end{equation}
Solutions for the two-mode entanglement of the ground state have been
presented in Refs. \cite{PRL,bec ground state ent}, for the case
of an initial state of $N$ atoms distributed evenly between the wells.}

In order to interpret these figures it should be kept in mind that
the attractive BEC case generates a nearly perfect planar quantum-squeezed
(PQS) state. This is close to an optimal minimization of the variance
sum, reducing phase-noise in both quadratures. The resulting state
is used in direct interferometry, with a single beam-splitter. On
the other hand, while the repulsive BEC case also produces a PQS state,
it has a different characteristic with one of the quadratures having
a phase-noise level reduced almost to the Heisenberg limit, while
the other phase quadrature is not at the Heisenberg limit. This state
is used in Mach-Zehnder interferometry, with an additional phase rotation
to generate the state with the optimal characteristics for phase-measurement.

\subsection{Entanglement criteria}

\begin{figure}
\begin{raggedright}
\includegraphics[width=0.72\columnwidth]{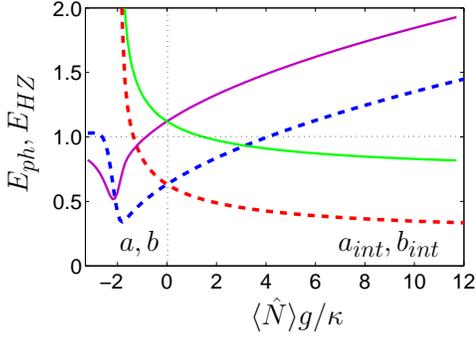}
\par\end{raggedright}

\caption{{\footnotesize (Color online) Entanglement of a ground state BEC including
Poissonian atom number fluctuations. Here $N=100$, for fixed $g/\kappa=10^{3}$.
(i) Entanglement between wells $a$ and $b$ is detected if $E_{ph}<1$
(blue and red dashed curves) or $E_{HZ}<1$ (equation (\ref{eq:hz})
(purple and green solid curves). Curves $a,b$ (purple solid and blue
dashed) are for two-well BEC modes $a,b$; curves $a_{int},b_{int}$
(green solid and red dashed) are for $a,b$ formed from BEC modes
$a_{i}$, $b{}_{i}$ input to the M-Z interferometer sequence depicted
in Figure 1.} \label{fig:entanglement-phase}}
\end{figure}

\textcolor{black}{We present solutions for the ground state of (\ref{hamgs-1-1}),
including number fluctuations as described in section \ref{sub:Experimental-number-fluctuations}.
The dashed curves in figure \ref{fig:entanglement-phase} show that
the normalized phase-entanglement criteria Eq. (\ref{eq:hznorm})
$E_{ph}<1$ detects the two-mode entanglement of the ground state
of a two-well BEC \cite{Esteve2008,Gross2010,Treutlein,PRL} in a
way that is almost immune to Poissonian number fluctuations. On the
other hand, the solid curves for $E_{HZ}$ reveal that the entanglement
detected via the un-normalized Hillery-Zubairy criterion $E_{HZ}<1$
is very easily destroyed by number fluctuations.}

\begin{figure}[t]
\begin{centering}
\includegraphics[scale=0.55]{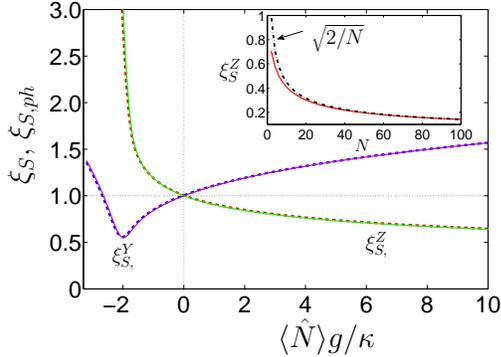}
\par\end{centering}

\caption{{\footnotesize (Color online) Spin squeezing and phase measurement
sensitivity parameters for ground state of a two-well BEC with $\langle\hat{N}\rangle=100$.
Plot of spin squeezing parameters $\xi_{S}^{Y}$, $\xi_{S}^{Z}$ (purple
and green solid curves) and the normalized parameters $\xi_{S,ph}^{Y}$,
$\xi_{S,ph}^{Z}$ (blue and red dashed curves), for Poissonian number
fluctuations. States with $\xi_{S}^{Y/Z},\xi_{S,ph}^{Y/Z}<1$ show
sub-shot noise enhanced phase sensitivity for measurements of small
rotations about a fixed phase. Inset shows the asymptotic $\sqrt{2/N}$
($(\Delta\phi)_{\pi/2}\sim O(\sqrt{2}/N)$) behavior with increasing
$N$ for fixed $g/\kappa=10^{3}$. The Heisenberg limit is $\xi_{S}^{Y/Z}\sim O(\sqrt{1/N})$
\cite{fisher}.}\textcolor{red}{\label{fig:Phase-measurement-sensitivity-1} }}
\end{figure}

\textcolor{black}{The figure \ref{fig:Phase-measurement-sensitivity-1}
shows that the normalized spin squeezing parameter also detects squeezing
and particle entanglement in a way that is insensitive to number fluctuations.
The solid curves plot the squeezing predicted for a fixed $N=100$,
where we recall that this parameter is not defined except in the case
of fixed $N$, but that according to Eq. (\ref{eq:spinsq}) will detect
both squeezing and entanglement among the $N$ particles when $\xi_{S}<1$.
The dashed curves plot the results for squeezing of the normalized
parameter, which requires $\xi_{S,norm}<1$ for detection of entanglement
but in the presence of arbitrary number $N$, to show perfect overlay.
Our proof justifies the normalization procedure used in recent experiments
that report spin squeezing and particle entanglement \cite{Esteve2008,Gross2010,Treutlein},
for a repulsive BEC with fluctuating total numbers. }

\textcolor{black}{The criteria Eq. (\ref{eq:hznorm}) and (\ref{eq:spinsqnorm})
enable an unambiguous detection of entanglement in the presence of
number fluctuations. We note that a similar immunity of the Peres
positive partial transpose PPT entanglement measure to loss was shown
in Ref. \cite{peres entlossbac}, and for other entanglement measures
to loss in Ref \cite{cavalentsteermulti}.}

\textcolor{black}{The figures include both attractive ($g<0$) and
repulsive ($g>0$) interactions. In the attractive case, the criteria
(\ref{eq:hz}-\ref{eq:hznorm}) for entanglement are satisfied when
applied directly to the modes of the two wells, then described by
$a$, $b$. In the repulsive case, an interferometric sequence is
necessary: two-well modes $a_{i},b_{i}$ are phase-shifted and placed
through a MZ beam-splitter\@. The reason for this is that while an
attractive BEC reduces fluctuations directly in the plane of $\hat{J}^{X,Y}$,
a repulsive BEC reduces fluctuations in a different plane, namely
in $\hat{J}^{X,Z}$. Without the additional input phase-shifter, the
phase-measurement is in the $Y-Z$ plane, where only one phase has
reduced fluctuations, as observed \cite{Gross2010}. }

\begin{figure}
\begin{raggedright}
\includegraphics[width=0.72\columnwidth]{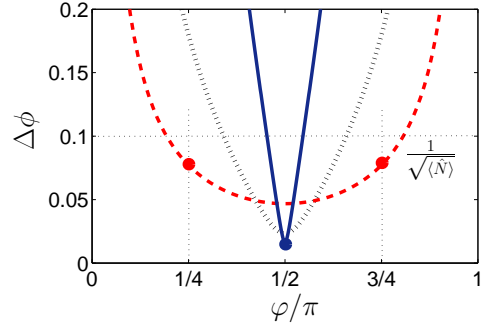}
\par\end{raggedright}

\caption{{\footnotesize (Color online) Measured phase uncertainty $\Delta\phi$
(\ref{eq:intexp}), for Poissonian fluctuations. Phase sensitivity
is better than shot noise level if $\Delta\phi<0.1=1/\sqrt{\langle\hat{N}\rangle}$.
The red dashed curve corresponds to the critical value of $\langle\hat{N}\rangle g/\kappa\simeq43.6$
which gives the lowest value of $E_{ph}$ for the repulsive regime.
Increasing $\langle\hat{N}\rangle g/\kappa$ to $10^{3}$ (dotted)
and $10^{4}$ (solid), improves the optimum phase uncertainty, but
the useful region of phase angles become narrower. Optimum sensitivity
is at $\varphi\pm\pi/2$ and is determined by the spin squeezing parameter
$\xi_{S,norm}$ . For detecting an unknown phase using only two orthogonal
measurements, the sensitivity over the quiet quadrant $\varphi/\pi=\frac{1}{2}\pm\frac{1}{4}$
becomes relevant.} \label{fig:entanglement-phase-1}}
\end{figure}

\subsection{\textcolor{black}{BEC interferometric phase measurement}}

\textcolor{black}{Now we turn to the question of using the BEC two-mode
states for the purpose of interferometric phase measurement. The spin
squeezing measure $\xi_{S}$ in the presence of fixed $N$, and the
normalized parameter $\xi_{S,norm}$ in the presence of fluctuating
numbers, give sensitivity for measurements of small rotations about
a fixed phase below the SQL when $\xi_{S}<1$ and $\xi_{S,norm}<1$
(Eq. (\ref{eq:sqnormintsql})). The figure \ref{fig:Phase-measurement-sensitivity-1}
reveals this spin squeezing to be predicted for a wide range of parameters
of the ground state solution. The inset shows the reduction in noise
to be near Heisenberg limited with the scaling $ $}{\footnotesize $\xi_{S}^{Y/Z}\sim O(\sqrt{2/N})$}\textcolor{black}{{}
evident. }

\textcolor{black}{The worse-case sensitivity of the interferometer
to an arbitrary angle defined within the two quiet quadrants of measurement
for an unknown phase $\phi$ is given by the $\eta_{ph}$ of Eq. (\ref{eq:phmeasuresubshot}),
which is minimized according to the phase entanglement measure $E_{ph}$
of figure \ref{fig:entanglement-phase-1}. The best scaling of $E_{ph}$
with $N$ is given as $J^{2/3}$, that of the $C_{J}$ coefficients,
and is achieved at the critical value of $Ng/\kappa$. This implies,
from the phase sensitivity measure $\eta_{ph}$, a sensitivity of
$(\Delta\phi)_{worse}\sim O(N^{-2/3})$.}

\textcolor{black}{For the proposed BEC interferometer, one can evaluate
the actual range of sensitivities for the unknown incoming phase $\phi$
using Eq (\ref{eq:intexp}) directly. The different range of phase-noise
reduction as a function of measured phase-angle and BEC interaction
strength is shown in figure \ref{fig:entanglement-phase-1}. The best
sensitivity is obtained at $\varphi=\pi/2$ and the value for $\Delta\phi$
is determined by the spin squeezing parameter Eq. (\ref{eq:spinsqnorm})
$\xi_{S,norm}$, which reduces to $\xi_{S}$ for fixed number $N$.
Where one measures an unknown phase $\phi$ using only two orthogonal
measurements ($\theta=0$ and $\theta'=\pi/2$). only the sensitivity
over the ``quiet'' quadrant indicated on the graph by the region
$\varphi/\pi=1/2\pm1/4$ becomes relevant. In this case, the worse
case sensitivities are at the edges ($\varphi=\pi/4$ and $\varphi=3\pi/4$)
and determined by the value of $\eta_{ph}$, Eq. (\ref{eq:sqnormintsql}).
This parameter is optimized by minimizing the two-mode entanglement
parameter $E_{ph}$, which reduces to the Hillery Zubairy entanglement
parameter when $N$ is fixed. This demonstrates the trade-off between
noise-reduction and range of measurable phase.}

\section{Conclusions}

In summary, we have introduced the normalized relative phase quadrature
operator $\tilde{J}\left(\phi\right)$ as the most direct operational
expression of how interferometric measurements give rise to phase
information. A corresponding phase-entanglement measure, as quantified
by $E_{ph}$, describes a useful physical resource for phase measurement.
This directly quantifies the measurement sensitivity increase above
the standard quantum limit, as $E_{ph}$ decreases towards a maximally
entangled state. We also introduce a normalized phase squeezing measure,
$\xi_{S,ph}^{Z/Y}$, which signifies entanglement between qubits or
particles. Both measures are normalized in terms of the total particle
number, using the generalized Moore-Penrose inverse method. We show
how it is possible to improve BEC phase measurements so that a range
of unknown phases have reduced phase-noise, rather than just one pre-selected
phase.

\section*{Appendix}

In this Appendix, we provide a detailed proof of the normalized entanglement
criterion, equation (\ref{eq:hznorm}) and the normalized spin-squeezing
criterion, (\ref{eq:spinsqnorm}).

\subsection*{Phase-entanglement criterion}

We wish to show that entanglement between modes $a$ and $b$ is confirmed
via an entanglement phase criterion, equation (\ref{eq:hznorm}).
First, we note the general result that if $\hat{N}$ commutes with
an arbitrary hermitian operator $\hat{O}$ having eigenvalues $j$,
we can introduce a limiting procedure to obtain a normalized mean
value, $\langle\tilde{O}\rangle=\langle\hat{O}\hat{N}^{+}\rangle$,
where $\hat{N}^{+}$ is the generalized Moore-Penrose \cite{Moore-Penrose}
inverse of the number operator $\hat{N}$, so that:
\begin{eqnarray}
\langle\hat{O}\hat{N}^{+}\rangle & = & \lim_{\epsilon\rightarrow0}\sum_{n,j}\frac{nj}{n^{2}+\epsilon}P(n,j)\nonumber \\
 & = & \sum_{n,j}n^{+}jP(n,j).
\end{eqnarray}
 Here $P(n,j)$ is the probability for simultaneous outcomes $n$,
$j$ for $\hat{N}$ and $\hat{O}$ respectively, and the eigenvalues
of the generalized inverse operator $\hat{N}^{+}$ are $n^{+}=n^{-1}$
for $n>0$, with $n^{+}=0$ for $n=0$. Hence, the expectation value
for the ratio becomes:
\[
\langle\tilde{O}\rangle=\sum_{n\geq0}\langle\tilde{O}\rangle_{n}P_{n},
\]
 where $\langle\hat{O}\rangle_{n}=\sum_{j}P(j|n)j$ , $P_{n}=\sum_{j}P(n,j)$,
$P(j|n)=P(n,j)/P_{n}$, and we define $\langle\tilde{O}\rangle_{n}=\langle\hat{O}\rangle_{n}n^{+}$.
Similarly, the corresponding variances, $\Delta^{2}\tilde{O}=\langle\tilde{O}^{2}\rangle-\langle\tilde{O}\rangle^{2}$
, can be expanded as:
\[
\Delta^{2}\tilde{O}=\sum_{n}\langle\tilde{O}^{2}\rangle_{n}P_{n}-\left[\sum_{n}\langle\tilde{O}\rangle_{n}P_{n}\right]^{2}.
\]
 However, we know from elementary variance properties that $\sum_{n}\langle\tilde{O}\rangle_{n}^{2}P_{n}\geq\left[\sum_{n}\langle\tilde{O}\rangle_{n}P_{n}\right]^{2}$,
hence we can write that:
\begin{eqnarray}
\Delta^{2}\tilde{O} & \geq & \sum_{n}\langle\tilde{O}^{2}\rangle_{n}P_{n}-\sum_{n}\langle\tilde{O}\rangle_{n}^{2}P_{n}\nonumber \\
 & = & \sum_{n}\left[\langle\tilde{O}^{2}\rangle_{n}-\langle\tilde{O}\rangle_{n}^{2}\right]P_{n}\nonumber \\
 & = & \sum_{n}\left[\Delta_{n}^{2}\hat{O}\right]\left(n^{+}\right)^{2}P_{n}.\label{eq:NormalizedVarianceInequality}
\end{eqnarray}
Here, as usual, we have defined $\Delta_{n}^{2}\hat{O}=\sum_{j}P(j|n)j^{2}-\langle\hat{O}\rangle_{n}^{2}$. 

Next, we apply this result to normalized spin variances, giving:
\begin{equation}
\bigl(\Delta\tilde{J}^{X}\bigr)^{2}+\bigl(\Delta\tilde{J}^{Y}\bigr)^{2}\geq\sum_{n}\left(n^{+}\right)^{2}P_{n}[\Delta_{n}^{2}\hat{J}^{X}+\Delta_{n}^{2}\hat{J}^{Y}],
\end{equation}
where we have used the definitions that:
\begin{eqnarray*}
\Delta^{2}\tilde{J}^{X} & = & \sum_{n}P_{n}\langle\tilde{J}^{X}{}^{2}\rangle_{n}-\langle\tilde{J}^{X}\rangle^{2}.
\end{eqnarray*}
Finally, if we assume the separability equation (\ref{eq:hzsep}),
we see that for a separable density matrix with a fixed total particle
number:$ $
\[
\Delta_{n}^{2}\hat{J}^{X}+\Delta_{n}^{2}\hat{J}^{Y}\geq n/2\,,
\]
which means that for normalized operators,
\begin{eqnarray*}
\bigl(\Delta\tilde{J}^{X}\bigr)^{2}+\bigl(\Delta\tilde{J}^{Y}\bigr)^{2} & \geq & \frac{1}{2}\sum_{n}n\left(n^{+}\right)^{2}P_{n}\\
 & = & \frac{1}{2}\sum_{n}n^{+}P_{n}\\
 & = & \frac{1}{2}\langle\hat{N}^{+}\rangle.
\end{eqnarray*}

Here we note that the generalized inverse has many of the properties
of the standard inverse, in particular that $\hat{N}\left(\hat{N}^{+}\right)^{2}=\hat{N}^{+}$,
and that of course no singularities occur with this criterion. When
this condition is violated, we must have an entangled state, which
leads to (\ref{eq:hznorm}).

\subsection*{Phase-squeezing criterion}

Next, we wish to show that entanglement between $N$ spin-$1/2$ systems,
where the number of systems can fluctuate, is confirmed by the normalized
spin squeezing criterion of equation (\ref{eq:spinsqnorm}). For $N$
spin $1/2$ separable systems, where $N$ is fixed and nonzero, one
finds that \cite{naturespinsqent}:
\begin{eqnarray*}
\bigl(\Delta\hat{J}^{Z}\bigr)^{2} & \geq & \frac{1}{N}[\langle\hat{J}^{X}\rangle^{2}+\langle\hat{J}^{Y}\rangle^{2}]\\
 & \geq & N^{+}\langle\hat{J}^{X}\rangle^{2}.
\end{eqnarray*}
The last expression, using the generalized inverse of $N$, holds
even when $N=0$. Thus, using the normalized variance inequality (\ref{eq:NormalizedVarianceInequality}),
we obtain:
\begin{eqnarray}
\Delta^{2}\tilde{J}^{Z} & \geq & \sum_{n}P_{n}\left(n^{+}\right)^{2}[\Delta_{n}^{2}\hat{J}^{Z}],\\
 & \geq & \sum_{n}P_{n}n^{+}\langle\hat{J}^{X}\hat{N}^{+}\rangle_{n}^{2}.
\end{eqnarray}
Therefore, using the result that $n^{2}n^{+}=n$, we se that:
\[
\langle\hat{N}\rangle\Delta^{2}\tilde{J}^{Z}\geq\{\sum_{n>0}P_{n}n^{2}n^{+}\}\sum_{n>0}P_{n}n^{+}\langle\hat{J}^{X}\hat{N}^{+}\rangle_{n}^{2}.
\]
Next, we use the Cauchy-Schwarz inequality: $\{\sum_{n>0}x_{n>0}^{2}\}\{\sum_{n>0}y_{n}^{2}\}\geq|\sum_{n>0}x_{n}y_{n}|^{2}$
where $x_{n}=\sqrt{nP_{n}}$ and $y_{n}=\sqrt{P_{n}n^{+}\langle\hat{J}^{X}\hat{N}^{+}\rangle_{n}}$,
and hence:
\begin{eqnarray*}
\langle\hat{N}\rangle\Delta^{2}\tilde{J}^{Z} & \geq & \left[\sum_{n>0}P_{n}\langle\hat{J}^{X}\hat{N}^{+}\rangle_{n}\right]^{2}\\
 & = & \left|\langle\hat{J}^{X}\hat{N}^{+}\rangle\right|^{2}=|\langle\tilde{J}^{X}\rangle|^{2}.
\end{eqnarray*}
This proves the phase-squeezing criterion, equation (\ref{eq:spinsqnorm}).
\begin{acknowledgments}
We wish to thank the Australian Research Council for funding via ACQAO
COE, Discovery and DECRA grants, as well as useful discussions with
M. Oberthaler, P. Treutlein and A. Sidorov. One of us (P.D.D.) wishes to thank the Aspen Centre for Physics for their generous hospitality.
\end{acknowledgments}

\end{document}